\documentstyle[prd,aps]{revtex}
\newcommand{\be}{\begin{equation}}
\newcommand{\ee}{\end{equation}}
\newcommand{\bea}{\begin{eqnarray}}
\newcommand{\eea}{\end{eqnarray}}

\newcommand{\bei}{\begin{itemize}}
\newcommand{\eei}{\end{itemize}}
\newcommand{\sla}[1]{#1\!\!\!\slash}

\newcommand{\rar}{\rightarrow}

\newcommand{\pp}{p^\prime}
\newcommand{\xp}{x^\prime}

\newcommand{\kp}{k^\prime}

\newcommand{\Mb}{M_{\Lambda_b}}

\newcommand{\MJ}{M_{J/\psi}}

\newcommand{\fb}{f_{\Lambda_b}}
\newcommand{\fL}{f_{\Lambda}}

\newcommand{\Lb}{\Lambda_b}
\newcommand{\La}{\Lambda}

\newcommand{\bp}{b^\prime}

\newcommand{\bo}{{\bf 0}}

\begin{document}
\author{
Chung-Hsien Chou\footnote{chouch@phys.sinica.edu.tw},
Hsien-Hung Shih\footnote{hhshih@phys.sinica.edu.tw},
Shih-Chang Lee\footnote{phsclee@ccvax.sinica.edu.tw}, and
Hsiang-nan Li\footnote{hnli@phys.sinica.edu.tw}
\\
{\small Institute of Physics, Academia Sinica, Taipei, Taiwan 115,
Republic of China} }

\title{$\Lambda_b \to \Lambda J/\psi$ decay in perturbative QCD}

\date{\today}

\maketitle

\begin{abstract}

We calculate the amplitudes involved in the heavy baryon nonleptonic
decay $\Lambda_b \rar \Lambda J/\psi$ using perturbative QCD
factorization theorem, which are expressed as convolutions of hard $b$
quark decay amplitudes with the $\Lambda_b$ baryon, $\Lambda$ baryon and
$J/\psi$ meson distribution amplitudes. It is found that nonfactorizable
contributions dominate over factorizable ones. Because of soft
cancellation in pairs of nonfactorizable diagrams, the
$\Lambda_b \rar \Lambda J/\psi$ decay is characterized by a large scale,
such that perturbative QCD is applicable. Employing the distribution
amplitudes determined in our previous works and from QCD sum rules, we
derive the branching ratio
$B(\Lambda_b \rar \Lambda J/\psi)=(1.7 \sim 5.3)\times 10^{-4}$ in
agreement with data. We predict an asymmetry parameter
$\alpha= -0.17 \sim -0.14$ associated with the anisotropic angular
distribution of the $\Lambda$ baryons produced in polarized $\Lambda_b$
baryon decays.

\end{abstract}

%\newpage

\section{INTRODUCTION}

Recently, we have developed perturbative QCD (PQCD) factorization
theorem for the semileptonic heavy baryon decays $\Lambda_b \rar p
l\bar{\nu}$ \cite{SLL1} and $\Lambda_b \rar \Lambda_c l\bar{\nu}$
\cite{SLL2,SLL3}. This theorem states that nonperturbative
dynamics involved in a high-energy QCD process can be factorized
into hadron distribution amplitudes, and the remaining piece is
calculable in perturbation theory. A distribution amplitude,
though not calculable, is universal. Once a distribution amplitude
is determined from experimental data of some processes, it can be
employed to make predictions for other processes involving the
same hadron. According to this theorem, the form factors involved
in the decay $\Lambda_b \rar \Lambda_c l\bar{\nu}$ have been
expressed as the convolutions of hard b quark decay amplitudes
with the universal $\Lambda_b$ and $\Lambda_c$ baryon distribution
amplitudes. It has been found that perturbative contributions to
the $\Lambda_b \rar \Lambda_c$ decays become more important at the
maximal recoil of the $\Lambda_c$ baryon with the velocity
transfer about 1.4. This observation indicates that PQCD is an
appropriate tool for analyses of two-body nonleptonic $\Lambda_b$ baryon
decays.

An essential feature of PQCD \cite{LY1,L1,CL,YL,WYL} is that it goes
beyond the conventional approach to exclusive nonleptonic heavy hadron
decays based on the factorization approximation (FA) \cite{BSW}. Both
factorizable and nonfactorizable contributions from various topologies
(emission and annihilation) can be evaluated systematically in PQCD.
Though nonfactorizable contributions are usually negligible, they become
dominant in the modes $B\to J/\psi K^{(*)}$ \cite{L2}, whose factorizable
contributions arise from internal $W$-emission with the small Wilson
coefficient $a_2$ defined below. This is the reason it is difficult to
accommodate the data of the ratios
$R=B(B\rar J/\psi K^*)/B(B\rar J/\psi K)$ and
$R_L=B(B\rar J/\psi K_L^*)/B(B\rar J/\psi K^*)$ simultaneously in FA.
However, these data can be explained in the PQCD formalism \cite{YL}.

In this paper we shall extend PQCD factorization theorem to the more
complicated baryon decay $\Lambda_b \rar \Lambda J/\psi$, and show that
nonfactorizable contributions also play an essential role. A simple
investigation indicates that each diagram for the
$\Lambda_b \rar \Lambda J/\psi$ decay is characterized by the scale
$\bar\Lambda$, where $\bar\Lambda=M_{\Lambda_b}-m_b$ is the mass
difference between the $\Lambda_b$ baryon and the $b$ quark. Factorizable
contributions, having this low typical scale, are suppressed by the
Wilson coefficient $a_2$. The dominant nonfactorizable
contributions, due to soft cancellation between a pair of diagrams,
are characterized by the higher scale
$\sqrt{\bar\Lambda \Mb[1-(\MJ/\Mb)^2]}$, with $\MJ$
being the $J/\psi$ meson mass. Therefore, the
$\Lambda_b \rar \Lambda J/\psi$ decay is a special heavy-to-heavy mode,
to which PQCD is applicable.

The $\Lambda_b$ baryon distribution amplitude has been chosen to satisfy
the experimental upper bound of the $\Lambda_b \rar \Lambda_c l\bar{\nu}$
branching ratio and the requirement of heavy quark symmetry \cite{SLL2}.
The $J/\psi$ meson distribution amplitudes have been determined from the
experimental data of the $B\to J/\psi K^{(*)}$ decays \cite{YL}. The
$\Lambda$ baryon distribution amplitudes have been derived from QCD sum
rules \cite{FZOZ,COZ}. We shall employ these distribution amplitudes,
because of their universality, to predict the branching ratio
$B(\Lambda_b \rar \Lambda J/\psi)$. We also consider another interesting
quantity, the asymmetry parameter $\alpha$ defined in Sec.~V, which is
related to the anisotropic angular distribution of the $\Lambda$ baryons
produced in polarized $\Lambda_b$ baryon decays.

The $\Lambda_b \rar \Lambda J/\psi$ decay has been discussed in
the quark model based on FA \cite{CT1}. Our predictions
\begin{eqnarray}
& &B(\Lambda_b \rar \Lambda J/\psi)= (1.7 \sim 5.3) \times
10^{-4} \;,
\nonumber\\
& &\alpha = -0.17 \sim -0.14 \;,
\end{eqnarray}
are consistent with those from FA. However, the theoretical bases
of the two approaches are quite different: nonfactorizable
contributions are neglected, and the Wilson coefficient $a_2$ is
treated as a free parameter in FA. We emphasize that there is no
free parameter in the PQCD calculation. The comparison of our
predictions with data will provide a justification of the PQCD
formalism for heavy baryon decays.

In Sec.~II we briefly review factorization theorem for exclusive
nonleptonic heavy hadron decays. The hadron distribution amplitudes and
the associated Sudakov factors are defined in Sec.~III. The factorization
formulas for the $\Lambda_b \rar \Lambda J/\psi$ decay amplitudes are
presented in Sec.~IV. Numerical results are discussed in Sec.~V. Section
VI is the conclusion. To deliver our ideas clearly, we shall leave all
technical calculations and complicated expressions to the Appendices.
Appendix A contains the useful integrals employed in this work and the
measures of phase-space integrations for different diagrams. The hard
amplitudes are summarized in Appendices B and C.

\section{FACTORIZATION THEOREM}

We briefly review PQCD factorization theorem for exclusive
nonleptonic heavy hadron decays \cite{CL}, and discuss its application
to the $\Lambda_b \rar \Lambda J/\psi$ mode. The effective Hamiltonian
responsible for this mode is expressed as
\bea
{\cal H}^{\rm eff} = \frac{G_F}{\sqrt{2}} V_{cb}V^*_{cs}
[\, C_1(\mu) O_1(\mu)+C_2(\mu) O_2(\mu)\,]\;,
\label{eff}
\eea
where $G_F$ is the Fermi coupling constant, $V$'s the
Cabibbo-Kobayashi-Maskawa matrix elements, $C_{1,2}(\mu)$ the
Wilson coefficients, and $\mu$ an arbitrary renormalization scale.
The four-fermion operators $O_{1,2}$ are written as
\bea
O_1 = (\bar{c}b) (\bar{s}c) \;,\;\;\;\;
O_2 = (\bar{s}b) (\bar{c}c) \;,
\eea
with $(\bar{q}_1 q_2)=\bar{q}_1\gamma_\mu(1-\gamma_5)q_2$ being the
$V-A$ current.

We shall have a careful look at the derivation of the above effective
Hamiltonian starting with eight-quark amplitudes (three quarks from the
$\Lambda_b$ baryon, three quarks from the $\Lambda$ baryon and two quarks
from the $J/\psi$ meson). The lowest-order diagrams contain one $W$ boson
for the weak decay of the $b$ quark and two hard gluons attaching the
spectator quarks. Nonleptonic $\Lambda_b$ baryon decays involve three
scales: the W boson mass $M_W$, at which the matching conditions of the
effective Hamiltonian are defined, the hard scale $t$ related to $\Mb$,
which reflects specific dynamics of different modes, and the
factorization scale of $O(\bar\Lambda)$ \cite{CL}. The factorization
scale is introduced to separate perturbative contributions from
nonperturbative contributions absorbed into hadron distribution
amplitudes $\phi(x,b)$, $x$ being a momentum fraction carried by one of
the valence quarks. In the PQCD formalism the factorization scale is
chosen as $1/b$, $b$ being the transverse extent of a hadron.

Radiative corrections generate various types of large logarithms, such
as $\alpha_s\ln(M_W/t)$ and $\alpha_s\ln(t b)$, which should be summed
by renormalization-group (RG) methods to give evolution factors. The
Wilson coefficients $C(t)$ from the summation of the first type of
logarithms correspond to the evolution from $M_W$ to $t$. The second type
of logarithms is summed to give the evolution $g(t,b)$ from $t$ to $1/b$,
which is governed by an anomalous dimension different from that of
$C(t)$. The difference arises from the loop corrections to the eight-quark
amplitudes for $g(t,b)$, and those to the four-quark amplitudes associated
with the $W$ boson exchange for $C(t)$ \cite{CLY}. There also exist
double logarithms $\alpha_s\ln^2(Pb)$, $P$ being the dominant light-cone
component of hadron momentum, which appear in radiative corrections to
hadron wave functions. These logarithms, from the overlap of collinear
and soft divergences, are treated by the resummation technique
\cite{CS,BS}. The result is a Sudakov exponential $\exp[-s(P,b)]$, which
decreases fast with $b$ and vanishes at $b=1/\Lambda_{\rm QCD}$,
$\Lambda_{\rm QCD}$ being the QCD scale. Since the Sudakov factor
supresses long-distance contributions from the large $b$ region
\cite{LS,KLS2}, the hard scale $t$, always larger than the factorization
scale $1/b$, does not go down to $\bar\Lambda$.

After summing logarithmic corrections, the hard amplitude $H(t)$ can be
calculated perturbatively by means of Feynman diagrams with on-shell
external quarks. The on-shellness in the present leading-power analysis
means that the virtualities of the external quarks are at most of
$O(\bar\Lambda/m_b)$. The $\Lambda_b \rar \Lambda J/\psi$ hard amplitude
contains both factorizable and nonfactorizable internal $W$-emission
contributions from the lowest-order diagrams in Fig.~1. Note that we do
not exhibit the diagrams, which are equivalent under exchange
of the $u$ and $d$ quarks. It will be observed that nonfactorizable
diagrams, especially those with both the charm quarks in the $J/\psi$
meson attached by the hard gluons, dominate. The vertex corrections to
the four-fermion operators, which are calculated in the improved QCD
factorization \cite{Ali,CT98,BBNS}, are of higher orders in the PQCD
approach and not considered in the present leading-order formalism.

A naive power counting for the $\Lambda_b \to \Lambda J/\psi$ decay
indicates that the important kinematic region corresponds to the parton
longitudinal and transverse momenta of $O(\bar\Lambda)$. This is the
reason each diagram for this mode is characterized by the scale
$\bar\Lambda$ as stated in the Introduction. The factorizable
contributions, characterized by this low scale, are suppressed by the
small Wilson coefficient $a_2=C_2+C_1/N_c$, $N_c$ being the number of
colors. The $\Lambda_b \to \Lambda J/\psi$ decay is then dominated by the
nonfactorizable contributions as in the $B\to J/\psi K^{(*)}$ decays.
Since the $J/\psi$ meson is a color-singlet object
with the two valence quarks moving parallelly, there exists soft
cancellation in a pair of nonfactorizable diagrams \cite{CKL} such
as Figs.~1(f) and 1(h). Hence, the parton momenta in the $\Lambda$
baryon become of $O(\Mb[1-(\MJ/\Mb)^2])$, and the $\Lambda_b \to
\Lambda J/\psi$ decay is characterized by the larger scale
$\sqrt{\bar\Lambda \Mb[1-(\MJ/\Mb)^2]}$. It then makes sense to
apply PQCD to this heavy-to-heavy decay mode.

At last, the $\Lambda_b\to \Lambda J/\psi$ decay amplitudes are expressed
as the convolution of the above factors,
\bea
C(t) \otimes H(t)\otimes g(t,b)\otimes \exp[-s(P,b)]
\otimes \phi(x,b)\;.
\label{ffc}
\eea
All the convolution factors, except $\phi(x,b)$, are calculable. The
hadron distribution amplitude $\phi(x,b)$, though not calculable, are
universal, since they absorb long-distance dynamics of a decay process,
which is insensitive to short-distance dynamics involved in the
$b$ quark decays. Based on universality, we employ the hadron
distribution amplitudes extracted from other experimental data or from QCD
sum rules to make predictions for the $\Lambda_b \rar \Lambda J/\psi$
decay. Note that the hard scale $t$ is a convolution variable, since the
derivation of Eq.~(\ref{ffc}) starts with the eight-quark, instead of
four-quark, amplitudes. Its precise expression should be determined by
diminishing next-to-leading-order corrections to the hard amplitude
$H(t)$. In this work we shall consider the different choices of $t$ as
the main source of the theoretical uncertainty.

\section{DISTRIBUTION AMPLITUDES}

We define kinematics of the initial and final hadrons as follows. The
$\Lb$ baryon is assumed to be at rest, and the $\Lambda$ baryon,
regarded as being massless, recoils in the minus direction. The momenta
$p$, $p'$ and $q=p-p'$ of the $\Lambda_b$ baryon, the $\Lambda$ baryon
and the $J/\psi$ meson, respectively, and the momenta of their valence
quarks are parametrized as
\begin{eqnarray}
& &p=(p^+,p^-,{\bf 0})=\frac{\Mb}{\sqrt{2}}(1,1,{\bf 0})\;,
\nonumber\\
& &k_1=(x_1 p^+,p^-,{\bf k}_{1T})\;,\;\;\;\;
k_2=(x_2 p^+,0,{\bf k}_{2T})\;,\;\;\;\;
k_3=(x_3 p^+,0,{\bf k}_{3T})\;,
\nonumber\\
& &\pp=(0,{\pp}^-,\bo)=(0,\rho p^-,\bo)\;,
\nonumber\\
& &\kp_1=(0,\xp_1 {\pp}^-,{\bf \kp}_{1T})\;,\;\;\;\;
\kp_2=(0,\xp_2 {\pp}^-,{\bf \kp}_{2T})\;,\;\;\;\;
\kp_3=(0,\xp_3 {\pp}^-,{\bf \kp}_{3T})\;,
\nonumber\\
& &q=(q^+,q^-,{\bf 0})=(p^+,r^2 p^-,{\bf 0})\;,
\nonumber\\
& &q_1=(y q^+,y q^-,{\bf q}_{T})\;,\;\;\;\;
q_2=((1-y) q^+,(1-y) q^-,-{\bf q}_{T})\;,
\end{eqnarray}
with the constants,
\be
\rho=1-r^2\;,\;\;\;\;r=\frac{M_{J/\psi}}{M_{\Lambda_b}}\;.
\ee
$k_1$ ($\kp_1$) is the $b$ ($s$) quark momentum, $x_i$ ($\xp_i$) are the
momentum fractions associated with the $\Lambda_b$ ($\Lambda$) baryon,
and ${\bf k}^{(\prime)}_T$ the corresponding transverse momenta,
satisfying $\sum_l{\bf k}^{(\prime)}_{lT}=0$.
$y$ is the momentum fraction and ${\bf q}_T$ the transverse momenta
carried by the charm quark in the $J/\psi$ meson.

The structure of the $\Lambda_b$ baryon wave function $Y_{\Lb}$
is simplified under the assumptions that the spin and orbital degrees of
freedom of the light quark system are decoupled, and that the $\Lambda_b$
baryon is in the ground state ($s$-wave). The wave function is given,
in the transverse momentum space, by \cite{RA}
\bea
(Y_{\Lb})_{\alpha\beta\gamma}(k_i,\mu) & = &
\frac{1}{2\sqrt{2}N_c} \int \prod_{l=2}^3 \frac{dw^-_ld{\bf w}_l}
{(2\pi)^3}
e^{ik_l\cdot w_l} \epsilon^{abc} \langle 0|T[b_\alpha^a(0)u_\beta^b(w_2)
d_\gamma^c(w_3)]|\Lambda_b(p)\rangle\;,
\nonumber \\
& = & \frac{f_{\Lambda_b}}{8\sqrt{2}N_c}
[(\sla{p}+M_{\Lambda_b})\gamma_5 C ]_{\beta\gamma}
[\Lambda_b(p)]_\alpha \Psi (k_i,\mu)\;,
\label{psii}
\eea
where $b$, $u$, and $d$ are the quark fields, $a$, $b$, and $c$ the color
indices, $\alpha$, $\beta$, and $\gamma$ the spinor indices, $C$ the
charge conjugation matrix, $\Lambda_b(p)$ the $\Lambda_b$ baryon spinor,
and $f_{\Lambda_b}$ the normalization constant.

The $\Lambda$ baryon wave functions are defined, in the
transverse momentum space, via
\begin{eqnarray}
(Y_{\La})_{\alpha\beta\gamma}(k'_i,\mu)&=&
\frac{1}{2\sqrt{2}N_c}\int \prod_{l=2}^{3}
\frac{d w_{l}^{+}d{\bf w}_l}
{(2\pi)^{3}}e^{\textstyle ik'_{l}\cdot w_{l}}\epsilon^{abc}
\langle 0|T[s_{\alpha}^{a}(0)u_{\beta}^{b}
(w_{2})d_{\gamma}^{c}(w_3)]|\La(p') \rangle\;,
\nonumber \\
&=&\frac{\fL}{8\sqrt{2}N_{c}}\bigg\{(\sla{p}^\prime C)_{\beta\gamma}
[\gamma_{5}\La(p')]_{\alpha}\Phi^V(\kp_{i},\mu)
+(\sla{p}^\prime \gamma_{5}C)_{\beta\gamma}[\La(p')]_{\alpha}
\Phi^A(\kp_{i},\mu)\bigg\}
\nonumber \\
& &-\frac{\fL^T}{8\sqrt{2}N_{c}}
(\sigma_{\mu\nu}{\pp}^\nu C)_{\beta\gamma}
[\gamma^{\mu}\gamma_{5}\La(p')]_{\alpha}\Phi^T(\kp_{i},\mu)\; ,
\label{4}
\end{eqnarray}
with the normalization constants $f_{\La}$ and $f^T_{\La}$, the
$\Lambda$ baryon spinor $\Lambda(p')$, and the definition
$\sigma_{\mu\nu}=[\gamma_\mu,\gamma_\nu]/2$. The wave functions
$\Phi^V$, $\Phi^A$, and $\Phi^T$ are associated with the different spin
structures of the three valence quarks in the $\La$ baryon.

The $J/\psi$ meson wave function is expressed as
\bea
(Y_{J/\psi})_{\alpha\beta}(q_i,\lambda,\mu) &=& \frac{1}{N_c} \int
\frac{d^4w}{(2\pi)^4}
e^{iq_1\cdot w} \langle 0|T[\bar{c}_\beta (0) c_\alpha (w)]
|J/\psi(q,\lambda)\rangle
\nonumber \\
&=& \frac{1}{\sqrt{2N_c}} [\sla{\epsilon}(q,\lambda)(\sla{q}
+M_{J/\psi})]_{\alpha\beta}\; \Pi(q_i,\mu) \;,
\eea
where the $J/\psi$ meson decay constant $f_{J/\psi}$ has been absorbed
into the wave function $\Pi$, and $\epsilon_\mu (q,\lambda)$ is the
polarization vector of the $J/\psi$ meson with the helicity $\lambda$.
It has been assumed that the $J/\psi$ meson
wave functions associated with longitudinal and transverse polarizations
possess the same form in the above expression. To ensure that the valence
charm quarks are close to the mass shell, $\Pi$ should have a sharp
peak at $y\sim 1/2$. The $J/\psi$ meson wave function has been
determined from the data of the $B\to J/\psi K^{(*)}$ decays \cite{YL}.
For the factorizable amplitudes, the momentum fraction $y$ is integrated
out, and only the decay constant $f_{J/\psi}$ is relevant.

When the transverse degrees of freedom of partons are taken into account,
the factorization of a QCD process should be constructed in the $b$
space \cite{BS}, with $b$ being the variable conjugate to $k_T$. This is
the reason $1/b$ serves as the factorization scale stated in Sec.~II.
Sudakov resummation of the double logarithms and RG summations of the
single logarithms contained in the above hadron wave functions lead to
\bea
\Psi(k_i^+,b_i,\mu) &=& {\rm exp}\left[ -\sum_{l=2}^3 s(w,k_l^+)
-3\int_{w}^\mu \frac{d\bar{\mu}}{\bar{\mu}}
\gamma (\alpha_s(\bar{\mu}))\right]
\psi(x_i) \;,
\nonumber\\
\Phi^j({\kp}_i^-,\bp_i,\mu) &=& \exp\left[ -\sum_{l=1}^3 s(w',{\kp}_l^-)
-3\int_{w'}^\mu \frac{d\bar{\mu}}{\bar{\mu}}
\gamma (\alpha_s(\bar{\mu}))\right]\phi^j(x_i) \;,
\nonumber\\
\Pi(q^+_i,b_q,\mu) &=& \exp\left[ -\sum_{l=1}^2 s(w_q,q_l^+)
-2\int_{w_q}^\mu \frac{d\bar{\mu}}{\bar{\mu}}
\gamma (\alpha_s(\bar{\mu}))\right]\pi(y) \;,
\label{wf}
\eea
with the superscript $j=V$, $A$ and $T$, and the quark anomalous dimension
$\gamma=-\alpha_s/\pi$. The Sudakov exponent $s$ for the $J/\psi$ meson
depends on the dominant component $q_l^+$. The factorization scales $w$,
$w'$ and $w_q$ are chosen as
\be
w=\min\left(\frac{1}{b_1},\frac{1}{b_2},\frac{1}{b_3}\right)\;,\;\;\;\;
w'=\min\left(\frac{1}{\bp_1},\frac{1}{\bp_2},\frac{1}{\bp_3}\right)\;,
\;\;\;\;
w_q=\frac{1}{b_q}\;,
\ee
with the variables,
\begin{equation}
b_1=|{\bf b}_2-{\bf b}_3|\;,\;\;\;\;
\bp_1=|{\bf \bp}_2-{\bf \bp}_3|\;.
\end{equation}
The initial conditions $\psi$, $\phi^j$ and $\pi$ of the Sudakov evolution
absorb nonperturbative dynamics below the factorization scales $w$,
$w'$ and $w_q$, respectively. Note that the Sudakov effect from the
$J/\psi$ meson is weak, since the distribution amplitude $\pi(y,\lambda)$
vanishes rapidly as $y\to 0, 1$.

The explicit expression of the Sudakov exponent $s$ is given by
\be
s(w,Q) = \int_{w}^Q \frac{dp}{p} \left[\ln \frac{Q}{p} A(\alpha_s (p))
+ B(\alpha_s (p)) \right] \;,
\label{s}
\ee
where the anomalous dimensions $A$ to two loops and $B$ to one loop are
\bea
A &=& C_F \frac{\alpha_s}{\pi}+ \left[ \frac{67}{9}-\frac{\pi^2}{3}
-\frac{10}{27}n_f+\frac{8}{3}\beta_0 \ln \frac{e^{\gamma_E}}{2} \right]
\left(\frac{\alpha_s}{\pi}\right)^2 \;,
\nonumber \\
B &=& \frac{2}{3} \frac{\alpha_s}{\pi} \ln \frac{e^{2\gamma_E-1}}{2} \;,
\eea
$C_F=\frac{4}{3}$ being a color factor, $n_f=4$ the flavor number,
and $\gamma_E$ the Euler constant. The one-loop running coupling constant,
\be
\frac{\alpha_s (\mu)}{\pi} = \frac{1}{\beta_0
\ln (\mu^2/\La^2_{\rm QCD})} \;,
\ee
with the coefficient $\beta_0=(33-2n_f)/12$, will be substituted into
Eq.~(\ref{s}).

\section{FACTORIZATION FORMULAS}

The $\Lb \rar J/\psi \La$ decay amplitude ${\cal M}$ is expressed as
\bea
{\cal M} = i\,\sum_{\lambda} \, {\cal M} (\lambda)\; ,
\eea
where the amplitude ${\cal M}(\lambda)$ can be decomposed into different
structures with the corresponding coefficients $A_1$, $A_2$, $B_1$,
and $B_2$:
\bea
{\cal M}(\lambda)= \bar{\Lambda}(p^{\prime})\left[ A_1\sla{\epsilon}
\gamma_5 + A_2\frac{p\cdot{\epsilon}}{\Mb}\gamma_5 + B_1\sla{\epsilon}
+ B_2 \frac{p\cdot{\epsilon}}{\Mb}\right]\Lambda_b(p)\;.
\eea
The general factorization formula for ${\cal M}(\lambda)$ is written as
\bea
{\cal M}(\lambda) &=&
\frac{G_F}{\sqrt{2}}V_{cb} V_{cs}^*  \int [{\cal D}x]\int [{\cal D}b]
({\bar{Y}}_{\La})_{\alpha^{\prime}\beta^{\prime}\gamma^{\prime}}
(\pp,\xp,b',\mu)
\nonumber \\
& &\times (Y_{J/\psi})^\dagger_{\rho\rho^\prime}(q,y,b_q,\lambda,\mu) \,
{\cal H}^{\alpha^\prime\beta^\prime\gamma^\prime\rho^\prime
\alpha\beta\gamma\rho}(x,\xp,y,b,\bp,b_q,\Mb,\mu)\,
(Y_{\Lb})_{\alpha\beta\gamma}(p,x,b,\mu)\;,
\label{da}
\eea
with the measure of the momentum fractions,
\bea
[{\cal D}x] = [dx][d\xp]dy\;,\;\;\;\;
[dx] = dx_1 dx_2 dx_3
\delta\left(1-\sum_{l=1}^3 x_l\right)\;,\;\;\;\;
[dx'] = dx'_1
dx'_2 dx'_3 \delta\left(1-\sum_{l=1}^3 x'_l\right)\;.
\eea
The measure of the transverse extents $[{\cal D}b]$ will be defined in
Appendix A.

The RG analysis of ${\cal H}$ leads to
\bea
{\cal H}(x,\xp,y,b,\bp,b_q,\Mb,\mu)
&=& \exp\left[ -n\int_\mu^t \frac{d\bar{\mu}}{\bar{\mu}}
\gamma (\alpha_s(\bar{\mu}))\right]
\nonumber\\
& &\times{\cal H} (x,\xp,y,b,\bp,b_q,\Mb,t) \;, \label{hp} \eea
with the integer $n=6$ for the factorizable diagrams and $n=8$ for
the nonfactorizable diagrams. The superscripts $\alpha^\prime$,
$\beta^\prime$, $\cdots$, have been suppressed. The argument $t$
in the initial condition of ${\cal H}$ implies that the Wilson
coefficient and the running coupling constants $\alpha_s$ for the
hard gluons are evaluated at $t$. This initial condition will be
calculated based on the lowest-order diagrams in Fig.~1. For the
purpose of presentation, we shall rewrite the initial hard
amplitude as ${\cal H}=a H_F \Omega$, where $a$ is the Wilson
coefficient, $H_F$ the numerator of ${\cal H}$ depending on the
spin structure of the three valence quarks in the $\Lambda$
baryon, and $\Omega$ the Fourier transformation of the denominator
of ${\cal H}$ from the $k_T$ space to the $b$ space.

Substituting Eqs.~(\ref{wf}) and (\ref{hp}) into Eq.~(\ref{da}),
we derive the factorization formulas,
\bea
F^{i} &=& 2 G_F
V_{cb}V^*_{cs}\sum_{j=V,A,T}\frac{\pi^2}{54\sqrt{3}} \fb \fL^{j}
\int [{\cal D}x] \int [{\cal D}b]^{i} [\alpha_s(t^i)]^2 a^i(t^{i})
\nonumber \\
& &\times\psi(x) {\phi^j}(\xp)\pi(y)\exp[-S^i] H^{ij}_F\Omega^{i}\;,
\label{fff}
\eea
where $F$ represents the coefficients $A_1$, $A_2$, $B_1$, and $B_2$,
the superscript $i$ labels the diagrams in Fig.~1, and
the superscript $j$ labels $V$, $A$, and $T$ associated with the
spin structures of the valence quarks in the $\Lambda$ baryon. The
factor 2 comes from the symmetry under the $u$-$d$ exchange. The
Wilson coefficients are $a^i=C_1/N_c$ for the nonfactorizable
diagrams with only one hard gluon attaching the $c$
quark or the $\bar c$ quark, and $a^i=a_2=C_2+C_1/N_c$ for the
factorizable diagrams and for the nonfactorizable diagrams with
two hard gluon attaching the two quarks of the $J/\psi$ meson. At the
characteristic scale of order $\sqrt{\bar\Lambda\Mb[1-(\MJ/\Mb)^2]}$,
$C_1/N_c$ is about few times larger than $a_2$. The Fourier integrals
employed in this work, and the resultant measures $[{\cal D}b]^{i}$
of the transverse extents for each diagram $i$ are
given in Appendix A. The explicit expressions of $\Omega^i$ are
presented in Appendix B. The functions $H_F^{ij}$ are
given in Appendix C. To derive the expressions of $H_F^{ij}$, we have
chosen $x_1=1$ and $x_2=x_3=0$, since $x_2$ and $x_3$, being of
$O(\bar\Lambda/m_b)$, are negligible in the current leading-power
analysis.

The main theoretical uncertainty of our predictions come from higher-order
corrections to the hard amplitudes, which can be reflected by the choice
of the hard scale $t^i$ \cite{CKL,KLS}. We shall consider the
following two choices:
\begin{eqnarray}
t^i &=& \max(t^i_1,t^i_2,w,w',w_q) \;,
\nonumber\\
t^i &=& \max\left(\frac{t^i_1+t^i_2}{2},w,w',w_q\right) \;,
\label{cht}
\end{eqnarray}
where the hard scales $t_1^i$ and $t_2^i$ associated with the two hard
gluons in each diagram $i$ are listed in Table I. The $\max$ in the above
expressions simply means that the hard scales should be larger than the
factorization scales. It is expected that the first choice will lead to
values lower than those from the second one. The theoretical uncertainty
can be reduced after next-to-leading-order corrections are included.

The exponents $S^i$ are given by
\bea
S^i &=& \sum_{l=2}^3 s(w,x_l p^+)+3\int_{w}^{t^i}
\frac{d\bar{\mu}}{\bar{\mu}} \gamma(\alpha_s(\bar{\mu}))
\nonumber \\
  & & +\sum_{l=1}^3 s(w',\xp_l {\pp}^-)+3\int_{w'}^{t^i}
\frac{d\bar{\mu}}{\bar{\mu}} \gamma(\alpha_s(\bar{\mu}))\;,
\\
S^i &=& \sum_{l=2}^3 s(w,x_l p^+)+3\int_{w}^{t^i}
\frac{d\bar{\mu}}{\bar{\mu}} \gamma(\alpha_s(\bar{\mu}))
\nonumber \\
  & & +\sum_{l=1}^3 s(w',\xp_l {\pp}^-)+3\int_{w'}^{t^i}
\frac{d\bar{\mu}}{\bar{\mu}} \gamma(\alpha_s(\bar{\mu}))
\nonumber \\
  & & +\sum_{l=1}^2 s(w_q,y_l q^+)+2\int_{w_q}^{t^i}
\frac{d\bar{\mu}}{\bar{\mu}} \gamma(\alpha_s(\bar{\mu}))\;,
\eea
for the factorizable and nonfactorizable diagrams, respectively. The
Sudakov exponential associated the $J/\psi$ meson distribution amplitude
appear only in the nonfactorizable diagrams. The evolution factors
described by $\gamma$ correspond to $g(t,b)$ introduced in Sec.~II.

\section{NUMERICAL RESULTS}

In this section we evaluate the factorization formulas in
Eq.~(\ref{fff}) numerically. For the $\Lb$ baryon distribution amplitude
$\psi$, we adopt the model proposed in \cite{Sch},
\be
\psi (x_1,x_2,x_3) = N x_1 x_2 x_3 \exp\left[ -\frac{\Mb^2}{2\beta^2 x_1}
-\frac{m_l^2}{2\beta^2 x_2} -\frac{m_l^2}{2\beta^2 x_3} \right] \;,
\ee
where the shape parameter $\beta=1.0$ GeV, and the mass of light degrees
of freedom in the $\Lb$ baryon, $m_l=0.3$ GeV, have been determined in
\cite{SLL2}. The normalization
\be
\int [dx] \psi (x_1,x_2,x_3) = 1 \;,
\ee
leads to the constant $N=6.67\times 10^{12}$. The constant
$f_{\Lambda_b}=2.71\times 10^{-3}$ GeV$^2$ and the above $\Lambda_b$
baryon distribution amplitude $\psi$ have been chosen to satisfy the
experimental upper bound of the $\Lambda_b\to\Lambda_cl{\bar\nu}$
branching ratio and the heavy quark symmetry \cite{SLL2}.

The $\Lambda$ baryon distribution amplitudes have been derived using QCD
sum rules \cite{FZOZ,COZ}. The asymmetric distribution in the momentum
fractions of the three quarks implies $SU(3)$ symmetry breaking
\cite{COZ}. In this work we adopt the models proposed in \cite{FZOZ},
\bea
\phi^V(x_1,x_2,x_3)&=& 42\,\phi_{as}(x_1,x_2,x_3) \left[
0.111(x_3^2-x_2^2) +0.093(x_2-x_3) \right] \;,
\nonumber\\
\phi^A(x_1,x_2,x_3)&=&-42\phi_{as}(x_1,x_2,x_3) \left[0.093
(x_2^2+x_3^2)+0.376x_1^2\right.
\nonumber\\
& &\left. -0.194x_2x_3-0.207x_1(x_2+x_3)\right]\;,
\nonumber\\
\phi^T(x_1,x_2,x_3)&=&
42\,\phi_{as}(x_1,x_2,x_3) \left[ 3.6(x_2^2-x_3^2) +0.32(x_2-x_3) \right]\;,
\eea
with the asymptotic distribution amplitude,
\be
\phi_{as}(x_1,x_2,x_3) = 120\; x_1 x_2 x_3 \;.
\ee
The normalization constants $\fL$ and $\fL^T$ are chosen as
\be
\fL \;=\;0.45\times 10^{-2}\; {\rm GeV}^2 \;, \;\;\;\;
\fL^T \;=\; 0.2\times 10^{-2}\; {\rm GeV}^2 \;.
\ee
It is easy to observe that the above $\Lambda$ baryon distribution
amplitudes satisfy the relations,
\bea
& &\phi^V(x_1,x_2,x_3)=-\phi^V(x_1,x_3,x_2)\;,
\nonumber \\
& &\phi^A(x_1,x_2,x_3)=\phi^A(x_1,x_3,x_2)\;,
\nonumber \\
& &\phi^T(x_1,x_2,x_3)=-\phi^T(x_1,x_3,x_2)\;,
\label{rel}
\eea
and the normalization,
\be
\int [dx] \phi^{V,A,T} (x_1,x_2,x_3) = 1 \;.
\ee

It is most likely that the two charm quarks in the $J/\psi$ meson carry
the equal fractional momenta, such that they are close to the mass shell.
Hence, the following $J/\psi$ meson distribution amplitudes with sharp
peaks at the momentum fraction $y=1/2$ have been proposed \cite{YL},
\bea
\pi(y) = \frac{5\sqrt{6}}{2}\, f_{J/\psi}\, y^2 (1-y)^2 \;,
\label{daj}
\eea
with the $J/\psi$ meson decay constant $f_{J/\psi}=390$ MeV
\cite{CM}. It has been found that Eq.~(\ref{daj}) gives
the branching ratios $B(B\rar J/\psi K^{(*)})$ in
agreement with experimental data \cite{YL}.
We have confirmed that the PQCD predictions are not sensitive to the
functional form of $\pi$, as long as it has a sharp peak
at $y\sim 1/2$.

Employing the parameters $G_F=1.16\times 10^{-5}$ GeV$^{-2}$,
$V_{cb}=0.04$ and $V_{cs}=0.975$, $\La_{\rm QCD}=0.2$ GeV, and the masses
$M_{\Lambda_b}=5.624$ GeV and $M_{J/\psi}=3.097$ GeV, we obtain the
coefficients $A_1$, $A_2$, $B_1$, and $B_2$. The contributions
from the transverse component $\phi^T$ of the $\Lambda$
baryon distribution amplitudes are listed in Table II. These
coefficients, dominated by the nonfactorizable amplitudes, are
mainly imaginary. The $\Lambda_b \rar \Lambda J/\psi$ decay rate is
written as
\bea
\Gamma = \frac{1}{8\pi} \frac{p_c}{\Mb^2} |{\cal M}|^2\;,
\eea
where $p_c$ is the magnitude of either of the final-state particle
momentum in the center-of-mass frame. Following the choices of
the hard scale $t$ in Eq.~(\ref{cht}), we derive the branching ratio,
\be
B(\Lambda_b\to\Lambda J/\psi)= (1.7 \sim 5.3)\times 10^{-4}\;,
\label{br}
\ee
for the $\Lambda_b$ baryon lifetime $\tau =(1.24 \pm 0.08)\times 10^{-12}$
s. Here we do not consider the minor uncertainty from $\tau$. The value in
Eq.~(\ref{br}) is consistent with the experimental data \cite{Data},
\be
B(\Lambda_b\to \Lambda J/\psi)= (4.7\pm 2.8)\times 10^{-4} \;.
\ee

We also calculate the asymmetry parameter $\alpha$ associated with the
anisotropic angular distribution of the $\Lambda$ baryons emitted in
polarized $\Lambda_b$ baryon decays:
\bea
{d\Gamma}\propto (1+\alpha \pp\cdot {\cal P}) \;,
\eea
${\cal P}$ being the $\Lambda_b$ baryon polarization. The explicit
expression of $\alpha$ is given by \cite{CT1}
\bea
\alpha&=&\frac{4M^2_{J/\psi}Re(S^*P_2)
+2E^2_{J/\psi}Re[(S+D)^*P_1]}{2(|S|^2+|P_2|^2)M^2_{J/\psi} +
(|S+D|^2+|P_1|^2)E^2_{J/\psi}} \;,
\eea
with the factors,
\bea
S &=& -A_1\;,
\nonumber \\
D&=&-\frac{p^2_c}{E_{J/\psi}(E_{\Lambda}+M_{\Lambda})}(A_1-A_2)\;,
\nonumber \\
P_1 &=&-\frac{p_c}{E_{J/\psi}}\left[{\frac{M_{\Lambda_b}
+M_{\Lambda}}{E_{\Lambda}+M_{\Lambda}}B_1+B_2}\right]\;,
\nonumber \\
P_2 &=& \frac{p_c}{E_{\Lambda}+M_{\Lambda}}B_1\;,
\eea
where $E_{J/\psi}$ ($E_\Lambda$) is the energy of the $J/\psi$ meson
($\Lambda$ baryon), and the $\Lambda$ baryon mass $M_\Lambda$ has been
set to zero for consistency. The result is, following Eq.~(\ref{cht}),
\begin{eqnarray}
\alpha=-0.17 \sim -0.14\;,
\end{eqnarray}
consistent with those derived from
the quark model based on FA \cite{FR,MO} shown in Table III. The
predictions for $\alpha$ are very stable with respect to the variation
of the hard scale $t$ and of the hadron distribution amplitudes.
Therefore, it serves as an ideal quantity to test the PQCD approach.

\section{CONCLUSION}

In this paper we have analyzed the nonleptonic heavy baryon decay
$\Lambda_b\to \Lambda J/\psi$ using the PQCD formalism. It is a special
heavy-to-heavy mode, to which PQCD is applicable, because of the soft
cancellation between a pair of nonfactorizable diagrams. We have shown
that nonfactorizable contributions dominate in this mode, similar to
the heavy meson cases $B\to J/\psi K^{(*)}$, since factorizable
contributions are suppressed by the small Wilson coefficient $a_2$.
After considering the theoretical uncertainty arising from higher-order
corrections, we have derived the branching ratio
$B(\Lambda_b\to \Lambda J/\psi)$ and the asymmetry parameter $\alpha$
associated with the anisotropic angular distribution of the $\Lambda$
baryons produced in the polarized $\Lambda_b$ baryon decays. The latter
quantity is stable with respect to the variation of the hadron
distribution amplitudes and to higher-order corrections. The comparison
of our predictions with the experimental data will provide a test of the
PQCD formalism.

The PQCD results are consistent with those derived from other approaches
based on FA as indicated in Table III. However, we emphasize that the
theoretical bases between PQCD and FA are very different. In PQCD, $a_2$
is a Wilson coefficient, which is small in a wide range of the energy
scale. To explain the experimental data, large nonfactorizable
contributions are necessary. In FA nonfactorizable contributions are
neglected. To account for the data, $a_2$ must be treated as a free
parameter, and the value $a_2\sim 0.23$ \cite{CT1} has been adopted.
Also, nonfactorizable contributions are imaginary and their strong phases
can be evaluated in PQCD, while the parameter $a_2$ is real (or with an
arbitrary phase) in FA. As stressed in \cite{KEK}, the imaginary
nonfactorizable amplitudes determine the relative phases of the various
$B\to D\pi$ decay modes, which are essential for explaining the recent
${\bar B}_d\to D^{(*)0}\pi^0$ data \cite{Belle,CLEO}.

\vskip 1.0cm

This work was supported by the National Science Council of R.O.C. under
Grant Nos. NSC-90-2112-M-001-077, NSC-90-2811-M-001-042 and
NSC-90-2112-M-001-038.

\appendix

\section{FOURIER INTEGRATIONS AND $b$ MEASURES}

We list below the Fourier integration formulas that have been employed
in the derivation of the hard amplitudes in the $b$ space, where $J_1$,
$N_1$, $K_0$ and $K_1$ are the Bessel functions, and $z_i$ the
Feynman parameters:
\bea
&  &\int d^2k \frac{e^{i {\bf k}\cdot {\bf b}}}{k^2+A} =
2\pi K_0(\sqrt{A}b)\;,  \;\;\;\; A \,>\, 0\;,
\\
& &\nonumber\\
& &\int d^2k \frac{e^{i {\bf k}\cdot {\bf b}}}{(k^2+A)(k^2+B)}
=\pi\int_0^1 dz\frac{K_1(\sqrt{Z_1}|b|)}{\sqrt{Z_1}}\;,
\;\;\;\; A,\,B\; >\; 0\;,
\\
& &\nonumber\\
& &\int d^2k_1d^2k_2 \frac{e^{i ({\bf k}_1\cdot {\bf b}_1
+{\bf k}_2\cdot {\bf b}_2)}}{(k_1^2+A)(k_2^2+B)[(k_1+k_2)^2+C]}
\nonumber \\
&= &\pi^2\int_0^1\frac{dz_1dz_2}{z_1(1-z_1)}
\frac{\sqrt{X_2}}{\sqrt{|Z_2|}} \Bigg\{K_1(\sqrt{X_2Z_2})\theta(Z_2)
\nonumber \\
& &+\frac{\pi}{2}\left[N_1(\sqrt{X_2|Z_2|}) -i
J_1(\sqrt{X_2|Z_2|})\right]\theta(-Z_2) \Bigg\}\;,
\nonumber\\
& &\;\;\;\; A\, >\, 0\;,\; {\rm and}\;B,\; C \; {\rm arbitrary}\;,
\\
& &\nonumber\\
& &\int d^2k_1d^2k_2d^2k_3 \frac{e^{i ({\bf k}_1\cdot {\bf b}_1
+{\bf k}_2\cdot {\bf b}_2 +{\bf k}_3\cdot {\bf b}_3)}}
{(k_1^2+A)(k_2^2+B)(k_3^2+C)[(k_1+k_2+k_3)^2+D]}
\nonumber  \\
&=&\pi^3\int_0^1\frac{dz_1dz_2dz_3}{z_1(1-z_1)z_2(1-z_2)}
\frac{\sqrt{X_3}}{\sqrt{|Z_3|}} \Bigg\{K_1(\sqrt{X_3Z_3})\theta(Z_3)
\nonumber \\
& &+\frac{\pi}{2}\left[N_1(\sqrt{X_3|Z_3|}) -i
J_1(\sqrt{X_3|Z_3|})\right]\theta(-Z_3) \Bigg\}\;,
\nonumber\\
& &\;\;\;\;A,\;B\;>\;0\;, \;{\rm and}\;C,\;D\;{\rm arbitrary}\;,
\eea
with the variables,
\bea
Z_1&=&A\;z+B\;(1-z)\;,
\\
Z_2 &=& A\;(1-z_2)+\frac{z_2}{z_1(1-z_1)}[B\;(1-z_1)+C\;z_1] \;,
\nonumber \\
X_2 &=&(b_1-z_1b_2)^2+\frac{z_1 (1-z_1)}{z_2}b_2^2 \;,
\\
Z_3 &=& A\;(1-z_3)+\frac{z_3}{z_2(1-z_2)}\left\{B\;(1-z_2)
+\frac{z_2}{z_1(1-z_1)}[C\;(1-z_1)+D\;z_1]\right\} \;,
\nonumber\\
X_3 &=& [b_1-b_2z_2-b_3z_2(1-z_1)]^2+\frac{z_2 (1-z_2)}{z_3}
(b_2-b_3 z_1)^2 \nonumber \\ & &+\frac{z_1 (1-z_1) z_2
(1-z_2)}{z_2 z_3}b_3^2 \;.
\eea

After performing the above integrations, we combine the $z$ measures
with the ordinary $b$ measures,
\begin{equation}
[db_l]=\frac{d^2b_l}{(2\pi)^2}\;,
\end{equation}
to form the special $b$ measures $[{\cal D}b]$ appearing in
Eq.~(\ref{fff}). $[{\cal D}b]^i$ for each diagram $i$ in Fig.~1 is
given by
\bea
[{\cal D}b]^{(a)}=[{\cal D}b]^{(j)}
=(2\pi)^3 [db_2][d\bp_2][d\bp_3] dz_1 dz_2 \;,
\eea
\bea
[{\cal D}b]^{(b)}=[{\cal D}b]^{(c)}=[{\cal D}b]^{(g)}=[{\cal D}b]^{(i)}
=[{\cal D}b]^{(m)}=[{\cal D}b]^{(n)}=[{\cal D}b]^{(o)}=[{\cal D}b]^{(p)}
=(2\pi)^4 [db_2][d\bp_2][d\bp_3][db_q]\;,
\eea
\bea
[{\cal D}b]^{(d)}=(2\pi)^3 [db_2][d\bp_2][d\bp_3] dz_1\;,
\eea
\bea
[{\cal D}b]^{(e)}=[{\cal D}b]^{(h)}
=(2\pi)^3 [d\bp_2][d\bp_3][db_q] dz_1 dz_2\;,
\eea
\bea
[{\cal D}b]^{(f)}=(2\pi)^3 [d\bp_2][d\bp_3][db_q] dz_1 dz_2 dz_3\;,
\eea
\bea
[{\cal D}b]^{(k)}=[{\cal D}b]^{(l)}=[{\cal D}b]^{(q)}=[{\cal D}b]^{(r)}
=(2\pi)^4 [db_2][db_3][d\bp_2][d\bp_3]\;.
\eea

\section{EXPRESSIONS OF $\Omega^i$}

Employing the integration formulas in Appendix A, we derive the hard
amplitudes in the $b$ space. The hard function $\Omega^i$ for each
diagram $i$ in Fig.~1, which arises from Fourier transformation of the
denominators of the internal particle propagators, is expressed as
\bea
\Omega^{(a)} &=& \frac{1}{4z_1(1-z_1)} \frac{\sqrt{B_a}}{\sqrt{Z_a}}
K_0\left(\sqrt{x_3\xp_3\rho}\Mb \bp_3\right)
K_1\left(\sqrt{Z_a B_a}\right)\;,
\eea
with
\bea
B_a &=& \frac{z_1 (1-z_1)}{z_2}(b_2-\bp_2)^2+[(1-z_1)b_2+z_1\bp_2]^2\;,
\nonumber \\
Z_a &=& x_2\xp_2\rho\Mb^2(1-z_2)+\frac{z_2}{z_1(1-z_1)}[(1-x_1)z_1
+x_3(1-\xp_2)(1-z_1)]\rho\Mb^2  \;,
\nonumber \\
\eea
\bea
\Omega^{(b)} &=&
K_0\left(\sqrt{x_2\xp_2\rho}\Mb (b_2+b_q)\right)
K_0\left(\sqrt{x_3\xp_3\rho}\Mb \bp_3\right)
\nonumber \\
& & \times K_0\left(\sqrt{x_3(1-\xp_2)\rho}\Mb |b_2-\bp_2|\right)
\bigg\{ K_0\left(\sqrt{Z_b}\Mb b_q\right) \;\theta (Z_b)
\nonumber \\
& &  +\frac{\pi}{2} \left[-N_0\left(\sqrt{|Z_b|}\Mb b_q\right)
+i J_0\left(\sqrt{|Z_b|}\Mb b_q\right)\right] \;\theta (-Z_b) \bigg\}\;,
\eea
with
\bea
Z_b = \frac{r^2}{4}-(1-x_2-y)[\xp_2\rho +(1-y)r^2]\;,
\eea
\bea
\Omega^{(c)} &=&
K_0\left(\sqrt{x_2\xp_2\rho}\Mb |b_2-b_q|\right)
K_0\left(\sqrt{x_3\xp_3\rho}\Mb \bp_3\right)
\nonumber \\
& &\times K_0\left(\sqrt{x_3(1-\xp_2)\rho}\Mb |b_2-\bp_2|\right)
\bigg\{ K_0\left(\sqrt{Z_c}\Mb b_q\right) \;\theta(Z_c)
\nonumber \\
& & +\frac{\pi}{2} \left[-N_0\left(\sqrt{|Z_c|}\Mb b_q\right)
+i J_0\left(\sqrt{|Z_c|}\Mb b_q\right)\right] \;\theta(-Z_c) \bigg\}\;,
\eea
with
\bea
Z_c = \frac{r^2}{4}+(x_2-y)[\xp_2\rho +(1-y)r^2]\;,
\eea
\bea
\Omega^{(d)} = \frac{1}{2\sqrt{Z_d}}
K_0\left(\sqrt{x_2\xp_2\rho}\Mb b_2\right)
K_0\left(\sqrt{x_3\xp_3\rho}\Mb \bp_3\right)
K_1\left(\sqrt{Z_d}\Mb |b_2-\bp_2|\right) \;,
\eea
with
\bea
Z_d = [x_3+\xp_2(1-x_3)\rho]\Mb^2z_1+
        x_3(1-\xp_2)\rho\Mb^2(1-z_1)\;,
\eea
\bea
\Omega^{(e)} &=& \frac{1}{4z_1(1-z_1)} \frac{\sqrt{B_e}}{\sqrt{|Z_e|}}
K_0\left(\sqrt{x_3\xp_3\rho}\Mb (\bp_3+b_q)\right)
\bigg\{ K_1\left(\sqrt{Z_eB_e}\right)\;\theta(Z_e)
\nonumber \\
& &  +\frac{\pi}{2} \left[N_1\left(\sqrt{|Z_e|B_e}\right)
-i J_1\left(\sqrt{|Z_e|B_e}\right)\right]\;\theta(-Z_e) \bigg\}\;,
\eea
with
\bea
B_e &=& \frac{z_1 (1-z_1)}{z_2}b_q^2+(\bp_2+z_1b_q)^2\;,
\nonumber \\
Z_e &=& x_2\xp_2\rho\Mb^2(1-z_2)+ \frac{z_2}{z_1(1-z_1)}
\Bigg\{ \left[\frac{r^2}{4}-(x_1-y)(1-\xp_1\rho -yr^2)\right]z_1
\nonumber \\
& & +\left[\frac{r^2}{4}-(1-x_3-y)(\xp_3\rho +(1-y)r^2)\right](1-z_1)
\Bigg\}\Mb^2\;,
\eea
\bea
\Omega^{(f)} &=& \frac{1}{8z_1(1-z_1)z_2(1-z_2)}
\frac{\sqrt{B_f}}{\sqrt{|Z_f|}}
\Bigg\{ K_1\left(\sqrt{Z_fB_f}\right)\;\theta(Z_f)
\nonumber \\
& & +\frac{\pi}{2} \left[N_1\left(\sqrt{|Z_f|B_f}\right)
-i J_1\left(\sqrt{|Z_f|B_f}\right)\right]\;\theta(-Z_f) \Bigg\}\;,
\eea
with
\bea
B_f &=& \frac{z_1(1-z_1)z_2(1-z_2)}{z_2z_3}b_q^2
+\frac{z_2(1-z_2)}{z_3}[\bp_3+(1-z_1)b_q]^2
\nonumber \\
& &
+\bigg\{\bp_2-b_qz_1-[\bp_3+(1-z_1)b_q]z_2\bigg\}^2\;,
\nonumber \\
Z_f &=& x_2\xp_2\rho\Mb^2(1-z_3)+\frac{z_3}{z_2(1-z_2)}
\Bigg\{ x_3\xp_3\rho\Mb^2(1-z_2)
\nonumber \\
& & +\frac{z_2}{z_1(1-z_1)}
 \left[ \left(\frac{r^2}{4}-(1-x_3-y_1)(\xp_3\rho +(1-y_1)r^2)\right)
 (1-z_1)\right.
\nonumber \\
& & \left.+\left(\frac{r^2}{4}+(x_2-y_1)(\xp_2\rho +y_1r^2)\right)z_1
\right]\Mb^2 \Bigg\}\;,
\eea
\bea
\Omega^{(g)} &=&
K_0\left(\sqrt{x_2\xp_2\rho}\Mb b_2\right)
K_0\left(\sqrt{x_3\xp_3\rho}\Mb (\bp_3+b_q)\right)
\nonumber \\
& & \times K_0\left(\sqrt{x_3+\xp_2(1-x_3)\rho}\Mb |b_2-\bp_2|\right)
\bigg\{ K_0\left(\sqrt{Z_g}\Mb b_q\right) \;\theta(Z_g)
\nonumber \\
& & +\frac{\pi}{2} \left[-N_0\left(\sqrt{|Z_g|}\Mb b_q\right)
  +i J_0\left(\sqrt{|Z_g|}\Mb b_q\right)\right] \;\theta(-Z_g) \bigg\}\;,
\eea
with
\bea
Z_g &=& \frac{r^2}{4}-(1-x_3-y_1)[\xp_3\rho +(1-y_1)r^2]\;,
\eea
\bea
\Omega^{(h)} &=& \frac{1}{4z_1(1-z_1)} \frac{\sqrt{B_h}}{\sqrt{|Z_h|}}
K_0\left(\sqrt{x_3\xp_3\rho}\Mb |b_3-b_q|\right)
\bigg\{ K_1\left(\sqrt{Z_hB_h}\right)\;\theta(Z_h)
\nonumber \\
& &  +\frac{\pi}{2} \left[N_1\left(\sqrt{|Z_h|B_h}\right)
-i J_1\left(\sqrt{|Z_h|B_h}\right)\right]\;\theta(-Z_h) \bigg\}\;,
\eea
with
\bea
B_h &=& \frac{z_1 (1-z_1)}{z_2}b_q^2+(\bp_2-z_1b_q)^2\;,
\nonumber \\
Z_h &=& x_2\xp_2\rho\Mb^2(1-z_2)+\frac{z_2}{z_1(1-z_1)}
\Bigg\{ \left[\frac{r^2}{4}+(1-x_1-y_1)(1-\xp_1\rho -(1-y_1)r^2)\right]z_1
\nonumber \\
& &
+\left[\frac{r^2}{4}+(x_3-y_1)(\xp_3\rho +y_1r^2)\right](1-z_1)
\Bigg\} \Mb^2\;,
\eea
\bea
\Omega^{(i)} &=&
K_0\left(\sqrt{x_2\xp_2\rho}\Mb b_2\right)
K_0\left(\sqrt{x_3\xp_3\rho}\Mb |\bp_3-b_q|\right)
\nonumber \\
& &\times
K_0\left(\sqrt{x_3+\xp_2(1-x_3)\rho}\Mb |b_2-\bp_2|\right)
\bigg\{ K_0\left(\sqrt{Z_i}\Mb b_q\right) \;\theta(Z_i)
\nonumber \\
& & +\frac{\pi}{2} \left[-N_0\left(\sqrt{|Z_i|}\Mb b_q\right)
+i J_0\left(\sqrt{|Z_i|}\Mb b_q\right)\right] \;\theta(-Z_i) \bigg\}\;,
\eea
with
\bea
Z_i &=& \frac{r^2}{4}+(x_3-y)(\xp_3\rho +yr^2)\;,
\eea
\bea
\Omega^{(j)} &=& \frac{1}{4z_1(1-z_1)} \frac{\sqrt{B_j}}{\sqrt{Z_j}}
K_0\left(\sqrt{x_2\xp_2\rho}\Mb b_2\right)
K_1\left(\sqrt{Z_jB_j}\right)\;,
\eea
with
\bea
B_j &=& \frac{z_1 (1-z_1)}{z_2}(b_2-\bp_2)^2+[\bp_3-z_1(b_2-\bp_2)]^2\;,
\nonumber \\
Z_j &=& \frac{z_2}{z_1(1-z_1)}\bigg\{(1-\xp_1)(1-z_1)\rho
+[x_3+\xp_2(1-x_3)\rho] z_1\bigg\}\Mb^2
\nonumber \\
& & +x_3\xp_3\rho\Mb^2(1-z_2)\;,
\eea
\bea
\Omega^{(k)} &=&
K_0\left(\sqrt{(1-x_1)(1-\xp_1)\rho}\Mb \bp_2\right)
K_0\left(\sqrt{x_3\xp_3\rho}\Mb |b_2-b_3|\right)
\nonumber \\
& & \times
K_0\left(\sqrt{(1-x_1)\xp_3\rho}\Mb |b_2-\bp_2-b_3+\bp_3|\right)
K_0\left(\sqrt{(1-x_1)\rho}\Mb |b_3-\bp_3|\right)\;,
\eea
\bea
\Omega^{(l)} &=&
K_0\left(\sqrt{(1-x_1)(1-\xp_1)\rho}\Mb |b_2-b_3+\bp_3|\right)
K_0\left(\sqrt{x_3\xp_3\rho}\Mb |\bp_2-\bp_3|\right)
\nonumber \\
& & K_0\left(\sqrt{x_3(1-\xp_1)\rho}\Mb |b_2-\bp_2-b_3+\bp_3|\right)
K_0\left(\sqrt{(1-x_1)\rho}\Mb |b_3-\bp_3|\right)\;,
\eea
\bea
\Omega^{(m)} &=&
K_0\left(\sqrt{(1-x_1)(1-\xp_1)\rho}\Mb (\bp_2+b_q)\right)
K_0\left(\sqrt{x_3\xp_3\rho}\Mb |b_2-\bp_3|\right)
\nonumber \\
& & \times K_0\left(\sqrt{(1-x_1)\xp_3\rho}\Mb |b_2-\bp_2|\right)
\bigg\{ K_0\left(\sqrt{Z_m}\Mb b_q\right)\theta(Z_m)
\nonumber \\
& &+\frac{\pi}{2} \left[ -N_0 \left(\sqrt{|Z_m|}\Mb b_q\right)
+i \,J_0 \left(\sqrt{|Z_m|}\Mb b_q\right) \right] \theta(-Z_m) \bigg\}\;,
\eea
with
\be
Z_m =\frac{r^2}{4}+(y_1-x_1)[1-y_1r^2-\xp_1\rho]\;,
\ee
\bea
\Omega^{(n)} &=&
K_0\left(\sqrt{(1-x_1)(1-\xp_1)\rho}\Mb (b_2+b_q)\right)
K_0\left(\sqrt{x_3\xp_3\rho}\Mb |\bp_2-\bp_3|\right)
\nonumber \\
& & \times
K_0\left(\sqrt{x_3(1-\xp_1)\rho}\Mb |b_2-\bp_2|\right)
\bigg\{ K_0\left(\sqrt{Z_n}\Mb b_q\right)\theta(Z_n)
\nonumber \\
& & +\frac{\pi}{2} \left[ -N_0 \left(\sqrt{|Z_n|}\Mb b_q\right)
+i \,J_0 \left(\sqrt{|Z_n|}\Mb b_q\right) \right] \theta(-Z_n) \bigg\}\;,
\eea
with
\be
Z_n = \frac{r^2}{4}+(y_1-x_1)[1-y_1r^2-\xp_1\rho]\;,
\ee
\bea
\Omega^{(o)} &=&
K_0\left(\sqrt{(1-x_1)(1-\xp_1)\rho}\Mb |\bp_2-b_q|\right)
K_0\left(\sqrt{x_3\xp_3\rho}\Mb |b_2-\bp_3|\right)
\nonumber \\
& & \times
K_0\left(\sqrt{(1-x_1)\xp_3\rho}\Mb |b_2-\bp_2|\right)
\bigg\{ K_0\left(\sqrt{Z_o}\Mb b_q\right)\theta(Z_o)
\nonumber \\
& & +\frac{\pi}{2} \left[ -N_0 \left(\sqrt{|Z_o|}\Mb b_q\right)
+i \,J_0 \left(\sqrt{|Z_o|}\Mb b_q\right) \right] \theta(-Z_o) \bigg\}\;,
\eea
with
\be
Z_o = \frac{r^2}{4}+(1-x_1-y_1)[1-(1-y_1)r^2-\xp_1\rho]\;,
\ee
\bea
\Omega^{(p)} &=&
K_0\left(\sqrt{(1-x_1)(1-\xp_1)\rho}\Mb |b_2-b_q|\right)
K_0\left(\sqrt{x_3\xp_3\rho}\Mb |\bp_2-\bp_3|\right)
\nonumber \\
& &
K_0\left(\sqrt{x_3(1-\xp_1)\rho}\Mb |b_2-\bp_2|\right)
\bigg\{ K_0\left(\sqrt{Z_p}\Mb b_q\right)\theta(Z_p)
\nonumber \\
& & +\frac{\pi}{2} \left[ -N_0 \left(\sqrt{|Z_p|}\Mb b_q\right)
+i \,J_0 \left(\sqrt{|Z_p|}\Mb b_q\right) \right] \theta(-Z_p) \bigg\}\;,
\eea
with
\be
Z_p =\frac{r^2}{4}+(1-x_1-y_1)[1-(1-y_1)r^2-\xp_1\rho]\;,
\ee
\bea
\Omega^{(q)} &=&
K_0\left(\sqrt{(1-x_1)(1-\xp_1)\rho}\Mb |\bp_2+b_3-\bp_3|\right)
K_0\left(\sqrt{x_3\xp_3\rho}\Mb |b_2-b_3|\right)
\nonumber \\
& & \times K_0\left(\sqrt{(1-x_1)\xp_3\rho}\Mb
|b_2-\bp_2-b_3+\bp_3|\right)
K_0\left(\sqrt{(1-x_1)\rho}\Mb |b_3-\bp_3|\right)\;,
\eea
\bea
\Omega^{(r)} &=&
K_0\left(\sqrt{(1-x_1)(1-\xp_1)\rho}\Mb b_2\right)
K_0\left(\sqrt{x_3\xp_3\rho}\Mb |\bp_2-\bp_3|\right)
\nonumber \\
& & \times K_0\left(\sqrt{x_3(1-\xp_1)\rho}\Mb
|b_2-\bp_2-b_3+\bp_3|\right)
K_0\left(\sqrt{(1-x_1)\rho}\Mb |b_3-\bp_3|\right) \;.
\eea

\section{EXPRESSIONS OF $H_F^{ij}$}

The part $H_F^{ij}$ of the hard amplitudes, that varies among the spin
structures of the three valence quarks in the $\Lambda$ baryon and among
the form factors $A_{1,2}$ and $B_{1,2}$, are gathered in the following
tables. Their derivation is similar to that presented in \cite{BLJ} for
the proton form factor. The contributions to $H_F^{ij}$ from
Figs.~1(a), 1(k), 1(l), 1(n), 1(p), and 1(r) vanish.

\vskip 0.5cm

\begin{center}
\begin{tabular}{|c|c|c|} \hline \hline
$F$ & $j$ & $H_F^{(b)j}$ \\ \hline
 & $V$ &
$-8r(1-r^2)^2\Mb^5(1-\xp_2)$
\\ \cline{2-3}
 & $A$ &
 $8r(1-r^2)^2\Mb^5(1-\xp_2)$
\\ \cline{2-3}
 \raisebox{3ex}[0pt]{$A_1$} & $T$ &
$24r(1-r^2)^2\Mb^5(1-\xp_2)$
\\ \cline{1-3}
 & $V$ &
$16 r (1-r^2)\Mb^5 (1-\xp_2)(1-2y)$
\\ \cline{2-3}
 & $A$ &
$-16 r (1-r^2)\Mb^5 (1-\xp_2)(1-2y)$
\\ \cline{2-3}
 \raisebox{3ex}[0pt]{$A_2$} & $T$ &
$-48 r (1-r^2)\Mb^5 (1-\xp_2)(1-2y)$
\\ \cline{1-3}
 & $V$ &
$8r(1-r^2)^2\Mb^5(1-\xp_2)$
\\ \cline{2-3}
 & $A$ &
$-8r(1-r^2)^2\Mb^5(1-\xp_2)$
\\ \cline{2-3}
 \raisebox{3ex}[0pt]{$B_1$} & $T$ &
$-24r(1-r^2)^2\Mb^5(1-\xp_2)$
\\ \cline{1-3}
 & $V$ &
$16 r (1-r^2)\Mb^5 (1-\xp_2)(1-2y)$
\\ \cline{2-3}
 & $A$ &
$-16 r (1-r^2)\Mb^5 (1-\xp_2)(1-2y)$
\\ \cline{2-3}
 \raisebox{3ex}[0pt]{$B_2$} & $T$ &
$-48 r (1-r^2)\Mb^5 (1-\xp_2)(1-2y)$
\\ \hline\hline
\end{tabular}
\end{center}

%%\newpage

\begin{center}
\begin{tabular}{|c|c|c|} \hline \hline
$F$ & $j$ & $H_F^{(c)j}$ \\ \hline
 & $V$ &
$16 r(1-r^2)^2\Mb^5 (1-\xp_2)y$
\\ \cline{2-3}
 & $A$ &
$-16 r(1-r^2)^2\Mb^5 (1-\xp_2)y$
\\ \cline{2-3}
 \raisebox{3ex}[0pt]{$A_1$} & $T$ &
$-48 r(1-r^2)^2\Mb^5 (1-\xp_2)y$
\\ \cline{1-3}
 & $V$ &
$ 0 $
\\ \cline{2-3}
 & $A$ &
$ 0 $
\\ \cline{2-3}
 \raisebox{3ex}[0pt]{$A_2$} & $T$ &
$ 0 $
\\ \cline{1-3}
 & $V$ &
$-16 r(1-r^2)^2\Mb^5 (1-\xp_2)y$
\\ \cline{2-3}
 & $A$ &
$16 r(1-r^2)^2\Mb^5 (1-\xp_2)y$
\\ \cline{2-3}
 \raisebox{3ex}[0pt]{$B_1$} & $T$ &
$48 r(1-r^2)^2\Mb^5 (1-\xp_2)y$
\\ \cline{1-3}
 & $V$ &
$ 0 $
\\ \cline{2-3}
 & $A$ &
$ 0 $
\\ \cline{2-3}
 \raisebox{3ex}[0pt]{$B_2$} & $T$ &
$ 0 $
\\ \hline\hline
\end{tabular}
\end{center}

\begin{center}
\begin{tabular}{|c|c|c|} \hline \hline
$F$ & $j$ & $H_F^{(d)j}$ \\ \hline
 & $V$ &
$16 r(1-r^2)^2\Mb^5 (1-\xp_2)$
\\ \cline{2-3}
 & $A$ &
$-16 r(1-r^2)^2\Mb^5 (1-\xp_2)$
\\ \cline{2-3}
 \raisebox{3ex}[0pt]{$A_1$} & $T$ &
$-48 r(1-r^2)^2\Mb^5 (1-\xp_2)$
\\ \cline{1-3}
 & $V$ &
$ 0 $
\\ \cline{2-3}
 & $A$ &
$ 0 $
\\ \cline{2-3}
 \raisebox{3ex}[0pt]{$A_2$} & $T$ &
$ 0 $
\\ \cline{1-3}
 & $V$ &
$-16 r(1-r^2)^2\Mb^5 (1-\xp_2)$
\\ \cline{2-3}
 & $A$ &
$16 r(1-r^2)^2\Mb^5 (1-\xp_2)$
\\ \cline{2-3}
 \raisebox{3ex}[0pt]{$B_1$} & $T$ &
$48 r(1-r^2)^2\Mb^5 (1-\xp_2)$
\\ \cline{1-3}
 & $V$ &
$ 0 $
\\ \cline{2-3}
 & $A$ &
$ 0 $
\\ \cline{2-3}
 \raisebox{3ex}[0pt]{$B_2$} & $T$ &
$ 0 $
\\ \hline\hline
\end{tabular}
\end{center}

%%\newpage

\begin{center}
\begin{tabular}{|c|c|p{10.5cm}|} \hline \hline
$F$ & $j$ & \quad \quad \quad \quad \quad \quad \quad \quad \quad
\quad $H_F^{(e)j}$ \\
\hline
 & $V$ &
$4 r(1-r^2)\Mb^5 (6+r^2-2\xp_1+2r^2\xp_1+2\xp_3-2r^2\xp_3
\newline
-12y-4r^2y+4\xp_1y-4r^2\xp_1y +4y^2+4r^2y^2)$
\\ \cline{2-3}
 & $A$ &
$-4 r(1-r^2)\Mb^5 (6+r^2-2\xp_1+2r^2\xp_1+2\xp_3-2r^2\xp_3
\newline
-12y-4r^2y+4\xp_1y-4r^2\xp_1y +4y^2+4r^2y^2)$
\\ \cline{2-3}
 \raisebox{3ex}[0pt]{$A_1$} & $T$ &
$-8 r(1-r^2)\Mb^5 (5+r^2-\xp_1+r^2\xp_1+\xp_3-r^2\xp_3-10y
\newline
-2r^2y+2\xp_1y-2r^2\xp_1y +4y^2+2r^2y^2)$
\\ \cline{1-3}
 & $V$ &
$16 r(1-r^2)\Mb^5 (1+\xp_1+\xp_3-4y +2y^2)$
\\ \cline{2-3}
 & $A$ &
$-16 r(1-r^2)\Mb^5 (1+\xp_1+\xp_3-4y+2r^2)$
\\ \cline{2-3}
 \raisebox{3ex}[0pt]{$A_2$} & $T$ &
$-16 r(1-r^2)\Mb^5 (3+\xp_1+\xp_3 -8y +4y^2)$
\\ \cline{1-3}
 & $V$ &
$-4 r(1-r^2)\Mb^5 (-2+r^2-2\xp_1+2r^2\xp_1+2\xp_3-2r^2\xp_3
\newline
-4r^2y+4y+4\xp_1y-4r^2\xp_1y -4y^2+4r^2y^2)$
\\ \cline{2-3}
 & $A$ &
$4 r(1-r^2)\Mb^5 (-2+r^2-2\xp_1+2r^2\xp_1+2\xp_3-2r^2\xp_3
\newline
-4r^2y+4y+4\xp_1y-4r^2\xp_1y -4y^2+4r^2y^2)$
\\ \cline{2-3}
 \raisebox{3ex}[0pt]{$B_1$} & $T$ &
$-8 r(1-r^2)\Mb^5 (3-r^2+\xp_1-r^2\xp_1-\xp_3+r^2\xp_3-6y
\newline
+2r^2y-2\xp_1y+2r^2\xp_1y +4y^2-2r^2y^2)$
\\ \cline{1-3}
 & $V$ &
$16 r\Mb^5 (1-3\xp_1+3r^2\xp_1+\xp_3-r^2\xp_3-2r^2y+4\xp_1y
\newline
-4r^2\xp_1y-2y^2+2r^2y^2)$
\\ \cline{2-3}
 & $A$ &
$-16 r\Mb^5 (1-3\xp_1+3r^2\xp_1+\xp_3-r^2\xp_3-2r^2y+4\xp_1y
\newline
-4r^2\xp_1y-2y^2+2r^2y^2)$
\\ \cline{2-3}
 \raisebox{3ex}[0pt]{$B_2$} & $T$ &
$16 r\Mb^5 (1+r^2+3\xp_1-3r^2\xp_1-\xp_3+r^2\xp_3-2r^2y-4y
\newline
-4\xp_1y +4r^2\xp_1y+4y^2)$
\\ \hline\hline
\end{tabular}
\end{center}

%%\newpage

\begin{center}
\begin{tabular}{|c|c|c|} \hline \hline
$F$ & $j$ &  $H_F^{(f)j}$ \\ \hline
 & $V$ &
$-16 r(1-r^2)\Mb^5 (1-y)^2$
\\ \cline{2-3}
 & $A$ &
$16 r(1-r^2)\Mb^5 (1-y)^2$
\\ \cline{2-3}
 \raisebox{3ex}[0pt]{$A_1$} & $T$ &
$8 r(1-r^2)\Mb^5 (3-5y+2r^2y+2\xp_3y-2r^2\xp_3y+2y^2 -2r^2y^2)$
\\ \cline{1-3}
 & $V$ &
$-32 r(1-r^2)\Mb^5 (1-y)^2$
\\ \cline{2-3}
 & $A$ &
$32 r(1-r^2)\Mb^5 (1-y)^2$
\\ \cline{2-3}
 \raisebox{3ex}[0pt]{$A_2$} & $T$ &
$16 r(1-r^2)\Mb^5 (3-2y)(1-y)$
\\ \cline{1-3}
 & $V$ &
$-16 r(1-r^2)\Mb^5 (1-y)^2$
\\ \cline{2-3}
 & $A$ &
$16 r(1-r^2)\Mb^5 (1-y)^2$
\\ \cline{2-3}
 \raisebox{3ex}[0pt]{$B_1$} & $T$ &
$-8 r(1-r^2)\Mb^5 (-3 +5y+2r^2y+2\xp_3y -2r^2\xp_3y-2y^2-2r^2y^2)$
\\ \cline{1-3}
 & $V$ &
$32 r\Mb^5 (1-y)(1-y-r^2y)$
\\ \cline{2-3}
 & $A$ &
$32 r\Mb^5 (1-y)(-1+y+r^2y)$
\\ \cline{2-3}
 \raisebox{3ex}[0pt]{$B_2$} & $T$ &
$16 r\Mb^5 (-3 +5y+5r^2y+4\xp_3y -4r^2\xp_3y-2y^2-6r^2y^2)$
\\ \hline\hline
\end{tabular}
\end{center}

%\newpage
\vskip 0.5cm

\begin{center}
\begin{tabular}{|c|c|c|} \hline \hline
$F$ & $j$ &  $H_F^{(g)j}$ \\ \hline
 & $V$ &
$-8 r(1-r^2)\Mb^5 (2+\xp_2-r^2\xp_2-2y-2\xp_2y +2r^2\xp_2y )$
\\ \cline{2-3}
 & $A$ &
$8 r(1-r^2)\Mb^5 (2+\xp_2-r^2\xp_2-2y-2\xp_2y +2r^2\xp_2y )$
\\ \cline{2-3}
 \raisebox{3ex}[0pt]{$A_1$} & $T$ &
$8 r(1-r^2)\Mb^5 (4+2r^2+\xp_2-r^2\xp_2+2\xp_3-2r^2\xp_3-4y-2r^2y
-2\xp_2y+2r^2\xp_2y )$
\\ \cline{1-3}
 & $V$ &
$16 r(1-r^2)\Mb^5 (-2+2y+\xp_2)$
\\ \cline{2-3}
 & $A$ &
$-16 r(1-r^2)\Mb^5 (-2+2y+\xp_2)$
\\ \cline{2-3}
 \raisebox{3ex}[0pt]{$A_2$} & $T$ &
$16 r(1-r^2)\Mb^5 (4-\xp_2-4y )$
\\ \cline{1-3}
 & $V$ &
$-8 r(1-r^2)\Mb^5 (2-\xp_2+r^2\xp_2-2y+2\xp_2y -2r^2\xp_2y )$
\\ \cline{2-3}
 & $A$ &
$ 8 r(1-r^2)\Mb^5 (2-\xp_2+r^2\xp_2-2y+2\xp_2y -2r^2\xp_2y )$
\\ \cline{2-3}
 \raisebox{3ex}[0pt]{$B_1$} & $T$ &
$-8 r(1-r^2)\Mb^5 (-4+2r^2+\xp_2-r^2\xp_2+2\xp_3-2r^2\xp_3+4y
-2r^2y-2\xp_2y+2r^2\xp_2y)$
\\ \cline{1-3}
 & $V$ &
$16 r\Mb^5 (2-3\xp_2+3r^2\xp_2-2y-2r^2y +4\xp_2y-4r^2\xp_2y) $
\\ \cline{2-3}
 & $A$ &
$-16 r\Mb^5 (2-3\xp_2+3r^2\xp_2-2y-2r^2y +4\xp_2y-4r^2\xp_2y) $
\\ \cline{2-3}
 \raisebox{3ex}[0pt]{$B_2$} & $T$ &
$16 r\Mb^5 (-4+2r^2+3\xp_2-3r^2\xp_2+4\xp_3-4r^2\xp_3+8y-4r^2y
-4\xp_2y+4r^2\xp_2y )$
\\ \hline\hline
\end{tabular}
\end{center}

%%\newpage

\begin{center}
\begin{tabular}{|c|c|c|} \hline \hline
$F$ & $j$ &  $H_F^{(h)j}$ \\ \hline
 & $V$ &
$-8 r(1-r^2)\Mb^5 (r^2-y-r^2y-2\xp_3y+2r^2\xp_3y -2r^2y^2 )$
\\ \cline{2-3}
 & $A$ &
$8 r(1-r^2)\Mb^5 (r^2-y-r^2y-2\xp_3y+2r^2\xp_3y -2r^2y^2 )$
\\ \cline{2-3}
 \raisebox{3ex}[0pt]{$A_1$} & $T$ &
$4 r(1-r^2)\Mb^5 (3r^2 -6y-6r^2y-4\xp_3y+4r^2\xp_3y+4y^2
\newline
-4r^2y^2)$
\\ \cline{1-3}
 & $V$ &
$16 r(1-r^2)\Mb^5 y$
\\ \cline{2-3}
 & $A$ &
$-16 r(1-r^2)\Mb^5 y$
\\ \cline{2-3}
 \raisebox{3ex}[0pt]{$A_2$} & $T$ &
$16 r(1-r^2)\Mb^5 y(-3+2y)$
\\ \cline{1-3}
 & $V$ &
$8 r(1-r^2)\Mb^5 (r^2 +y-r^2y-2\xp_3y+2r^2\xp_3y-2r^2y^2 )$
\\ \cline{2-3}
 & $A$ &
$-8 r(1-r^2)\Mb^5 (r^2 +y-r^2y-2\xp_3y+2r^2\xp_3y-2r^2y^2 )$
\\ \cline{2-3}
 \raisebox{3ex}[0pt]{$B_1$} & $T$ &
$-4 r(1-r^2)\Mb^5 (3r^2 +6y-6r^2y-4\xp_3y+4r^2\xp_3y
\newline
-4y^2-4r^2y^2)$
\\ \cline{1-3}
 & $V$ &
$-16 r(1-r^2)\Mb^5 y$
\\ \cline{2-3}
 & $A$ &
$16 r(1-r^2)\Mb^5 y$
\\ \cline{2-3}
 \raisebox{3ex}[0pt]{$B_2$} & $T$ &
$-16 r(1-r^2)\Mb^5 y(-3+2y)$
\\ \hline\hline
\end{tabular}
\end{center}

%\newpage
\vskip 0.5cm

\begin{center}
\begin{tabular}{|c|c|c|} \hline \hline
$F$ & $j$ &  $H_F^{(i)j}$ \\ \hline
 & $V$ &
$8 r(1-r^2)\Mb^5 (1+r^2-2\xp_3+2r^2\xp_3 -2r^2y)$
\\ \cline{2-3}
 & $A$ &
$-8 r(1-r^2)\Mb^5 (1+r^2-2\xp_3+2r^2\xp_3 -2r^2y)$
\\ \cline{2-3}
 \raisebox{3ex}[0pt]{$A_1$} & $T$ &
$-8 r(1-r^2)\Mb^5 (3+3r^2-4\xp_3+4r^2\xp_3 -2y-4r^2y)$
\\ \cline{1-3}
 & $V$ &
$16 r(1-r^2)\Mb^5 (1-2\xp_2y )$
\\ \cline{2-3}
 & $A$ &
$-16 r(1-r^2)\Mb^5 (1-2\xp_2y )$
\\ \cline{2-3}
 \raisebox{3ex}[0pt]{$A_2$} & $T$ &
$16 r(1-r^2)\Mb^5 (-3+2y+2\xp_2y )$
\\ \cline{1-3}
 & $V$ &
$8 r(1-r^2)\Mb^5 (1-r^2+2\xp_3-2r^2\xp_3 +2r^2y)$
\\ \cline{2-3}
 & $A$ &
$-8 r(1-r^2)\Mb^5 (1-r^2+2\xp_3-2r^2\xp_3 +2r^2y)$
\\ \cline{2-3}
 \raisebox{3ex}[0pt]{$B_1$} & $T$ &
$-8 r(1-r^2)\Mb^5 (3-3r^2+4\xp_3-4r^2\xp_3 -2y+4r^2y)$
\\ \cline{1-3}
 & $V$ &
$-16 r(1-r^2)\Mb^5 (1 -2\xp_2y)$
\\ \cline{2-3}
 & $A$ &
$16 r(1-r^2)\Mb^5 (1 -2\xp_2y)$
\\ \cline{2-3}
 \raisebox{3ex}[0pt]{$B_2$} & $T$ &
$16 r(1-r^2)\Mb^5 (3 -2y-2\xp_2y)$
\\ \hline\hline
\end{tabular}
\end{center}

%%\newpage

\begin{center}
\begin{tabular}{|c|c|c|} \hline \hline
$F$ & $j$ &  $H_F^{(j)j}$ \\ \hline
 & $V$ &
 $16 r(1-r^2)\Mb^5 (1-\xp_2+r^2\xp_2)$
\\ \cline{2-3}
 & $A$ &
 $-16 r(1-r^2)\Mb^5 (1-\xp_2+r^2\xp_2)$
\\ \cline{2-3}
 \raisebox{3ex}[0pt]{$A_1$} & $T$ &
 $-16 r(1-r^2)\Mb^5 (3-\xp_2+r^2\xp_2)$
\\ \cline{1-3}
 & $V$ &
 $32 r(1-r^2)\Mb^5 (1-\xp_2)$
\\ \cline{2-3}
 & $A$ &
$-32 r(1-r^2)\Mb^5 (1-\xp_2)$
\\ \cline{2-3}
 \raisebox{3ex}[0pt]{$A_2$} & $T$ &
$32 r(1-r^2)\Mb^5 (-2+\xp_2)$
\\ \cline{1-3}
 & $V$ &
 $16 r(1-r^2)\Mb^5 (1+\xp_2-r^2\xp_2)$
\\ \cline{2-3}
 & $A$ &
  $-16 r(1-r^2)\Mb^5 (1+\xp_2-r^2\xp_2)$
\\ \cline{2-3}
 \raisebox{3ex}[0pt]{$B_1$} & $T$ &
 $-16 r(1-r^2)\Mb^5 (1+\xp_2-r^2\xp_2)$
\\ \cline{1-3}
 & $V$ &
 $-32 r(1-r^2)\Mb^5 (1+\xp_2)$
\\ \cline{2-3}
 & $A$ &
 $32 r(1-r^2)\Mb^5 (1+\xp_2)$
\\ \cline{2-3}
 \raisebox{3ex}[0pt]{$B_2$} & $T$ &
 $32 r(1-r^2)\Mb^5 (2\xp_1+\xp_2)$
\\ \hline\hline
\end{tabular}
\end{center}

%\newpage

%%\newpage

\begin{center}
\begin{tabular}{|c|c|c|} \hline \hline
$F$ & $j$ &  $H_F^{(m)j}$ \\ \hline
 & $V$ & $-8 r(1-r^2)^2\Mb^5 \xp_3 $
\\ \cline{2-3}
 & $A$ &
 $8 r(1-r^2)^2\Mb^5 \xp_3 $
\\ \cline{2-3}
 \raisebox{3ex}[0pt]{$A_1$} & $T$ &
 $-8 r(1-r^2)^2\Mb^5
 \xp_3$
\\ \cline{1-3}
 & $V$ &
 $16 r (1-r^2)\Mb^5( 1-2y)\xp_3$
\\ \cline{2-3}
 & $A$ &
 $-16 r (1-r^2)\Mb^5( 1-2y)\xp_3$
\\ \cline{2-3}
 \raisebox{3ex}[0pt]{$A_2$} & $T$ &
 $16 r (1-r^2)\Mb^5(1-2y)\xp_3$
\\ \cline{1-3}
 & $V$ & $8 r(1-r^2)^2\Mb^5 \xp_3 $
\\ \cline{2-3}
 & $A$ &
 $-8 r(1-r^2)^2\Mb^5 \xp_3 $
\\ \cline{2-3}
 \raisebox{3ex}[0pt]{$B_1$} & $T$ &
 $8 r(1-r^2)^2\Mb^5
 \xp_3$
\\ \cline{1-3}
 & $V$ &  $16 r (1-r^2)\Mb^5(1-2y)\xp_3$
\\ \cline{2-3}
 & $A$ &
$-16 r (1-r^2)\Mb^5(1-2y)\xp_3$
\\ \cline{2-3}
 \raisebox{3ex}[0pt]{$B_2$} & $T$ &
 $16 r (1-r^2)\Mb^5
(1-2y)\xp_3$
\\ \hline\hline
\end{tabular}
\end{center}

\begin{center}
\begin{tabular}{|c|c|c|} \hline \hline
$F$ & $j$ &  $H_F^{(o)j}$ \\ \hline
 & $V$ &  $16 r(1-r^2)^2\Mb^5 y \xp_3$
\\ \cline{2-3}
 & $A$ &
$-16 r(1-r^2)^2\Mb^5 y \xp_3$
\\ \cline{2-3}
 \raisebox{3ex}[0pt]{$A_1$} & $T$ &
$16 r(1-r^2)^2\Mb^5 y\xp_3$
\\ \cline{1-3}
 & $V$ &  $0$
\\ \cline{2-3}
 & $A$ & $0$
\\ \cline{2-3}
 \raisebox{3ex}[0pt]{$A_2$} & $T$ &
 $\,\,\,\,0$
\\ \cline{1-3}
 & $V$ & $-16 r(1-r^2)^2\Mb^5 y\xp_3$
\\ \cline{2-3}
 & $A$ &
$16 r(1-r^2)^2\Mb^5 y\xp_3$
\\ \cline{2-3}
 \raisebox{3ex}[0pt]{$B_1$} & $T$ &
$-16 r(1-r^2)^2\Mb^5 y\xp_3$
\\ \cline{1-3}
 & $V$ &  $0$
\\ \cline{2-3}
 & $A$ &
  $0$
\\ \cline{2-3}
 \raisebox{3ex}[0pt]{$B_2$} & $T$ &
 $\,\,\,\,0$
\\ \hline\hline
\end{tabular}
\end{center}

\begin{center}
\begin{tabular}{|c|c|c|} \hline \hline
$F$ & $j$ & $H_F^{(q)j}$ \\ \hline
 & $V$ &   $16 r(1-r^2)^2\Mb^5 \xp_3$
\\ \cline{2-3}
 & $A$ &
$-16 r(1-r^2)^2\Mb^5 \xp_3$
\\ \cline{2-3}
 \raisebox{3ex}[0pt]{$A_1$} & $T$ &
$16 r(1-r^2)^2\Mb^5 \xp_3$
\\ \cline{1-3}
 & $V$ &  $0$
\\ \cline{2-3}
 & $A$ &  $0$
\\ \cline{2-3}
 \raisebox{3ex}[0pt]{$A_2$} & $T$ &
 $0$
\\ \cline{1-3}
 & $V$ & $-16 r(1-r^2)^2\Mb^5 \xp_3$
\\ \cline{2-3}
 & $A$ &
 $16 r(1-r^2)^2\Mb^5 \xp_3$
\\ \cline{2-3}
 \raisebox{3ex}[0pt]{$B_1$} & $T$ &
$-16 r(1-r^2)^2\Mb^5 \xp_3$
\\ \cline{1-3}
 & $V$ &  $0$
\\ \cline{2-3}
 & $A$ &  $0$
\\ \cline{2-3}
 \raisebox{3ex}[0pt]{$B_2$} & $T$ &
  $0$
\\ \hline\hline
\end{tabular}
\end{center}

%\newpage

\newpage
\vskip 1.0cm

Table I: The hard scales $t_1^i$ and $t_2^i$ for each
diagram $i$ in Fig.~1.
\vskip 0.5cm

\begin{tabular}{cll}
\hline
i & $t_1^i$ & $t_2^i$
\\ \hline
(a) & max( $\sqrt{x_2 \xp_2\rho}\Mb$, $\frac{1}{|b_2|}$ )
    & max( $\sqrt{x_3 \xp_3\rho}\Mb$, $\frac{1}{|\bp_3|}$ )
\\
(b) & max( $\sqrt{x_2 \xp_2\rho}\Mb$, $\frac{1}{|b_2+b_q|}$ )
    & max( $\sqrt{x_3 \xp_3\rho}\Mb$, $\frac{1}{|\bp_3|}$ )
\\
(c) & max( $\sqrt{x_2 \xp_2\rho}\Mb$, $\frac{1}{|b_2-b_q|}$ )
    & max( $\sqrt{x_3 \xp_3\rho}\Mb$, $\frac{1}{|\bp_3|}$ )
\\
(d) & max( $\sqrt{x_2 \xp_2\rho}\Mb$, $\frac{1}{|b_2|}$ )
    & max( $\sqrt{x_3 \xp_3\rho}\Mb$, $\frac{1}{|\bp_3|}$ )
\\
(e) & max( $\sqrt{x_2 \xp_2\rho}\Mb$, $\frac{1}{|b_2|}$ )
    & max( $\sqrt{x_3 \xp_3\rho}\Mb$, $\frac{1}{|\bp_3+b_q|}$ )
\\
(f) & max( $\sqrt{x_2 \xp_2\rho}\Mb$, $\frac{1}{|b_2|}$ )
    & max( $\sqrt{x_3 \xp_3\rho}\Mb$, $\frac{1}{|\bp_3+b_q|}$ )
\\
(g) & max( $\sqrt{x_2 \xp_2\rho}\Mb$, $\frac{1}{|b_2|}$ )
    & max( $\sqrt{x_3 \xp_3\rho}\Mb$, $\frac{1}{|\bp_3+b_q|}$)
\\
(h) & max( $\sqrt{x_2 \xp_2\rho}\Mb$, $\frac{1}{|b_2|}$ )
    & max( $\sqrt{x_3 \xp_3\rho}\Mb$, $\frac{1}{|\bp_3-b_q|}$ )
\\
(i) & max( $\sqrt{x_2 \xp_2\rho}\Mb$, $\frac{1}{|b_2|}$ )
    & max( $\sqrt{x_3 \xp_3\rho}\Mb$, $\frac{1}{|\bp_3-b_q|}$ )
\\
(j) & max( $\sqrt{x_2 \xp_2\rho}\Mb$, $\frac{1}{|b_2|}$ )
    & max( $\sqrt{x_3 \xp_3\rho}\Mb$, $\frac{1}{|\bp_3|}$ )
\\
(k) & max( $\sqrt{(1-x_1)(1-\xp_1)\rho}\Mb$, $\frac{1}{|b_2|}$ )
    & max( $\sqrt{x_3 \xp_3\rho}\Mb$, $\frac{1}{|b_1|}$ )
\\
(l) & max( $\sqrt{(1-x_1)(1-\xp_1)\rho}\Mb$, $\frac{1}{|b_2-b_3+\bp_3|}$)
    & max( $\sqrt{x_3 \xp_3\rho}\Mb$, $\frac{1}{|\bp_1|}$ )
\\
(m) & max( $\sqrt{(1-x_1)(1-\xp_1)\rho}\Mb$, $\frac{1}{|\bp_2+b_q|}$ )
    & max( $\sqrt{x_3 \xp_3\rho}\Mb$, $\frac{1}{|b_2-\bp_3|}$ )
\\
(n) & max( $\sqrt{(1-x_1)(1-\xp_1)\rho}\Mb$, $\frac{1}{|b_2+b_q|}$ )
    & max( $\sqrt{x_3 \xp_3\rho}\Mb$, $\frac{1}{|\bp_1|}$ )
\\
(o) & max( $\sqrt{(1-x_1)(1-\xp_1)\rho}\Mb$, $\frac{1}{|\bp_2-b_q|}$ )
    & max( $\sqrt{x_3 \xp_3\rho}\Mb$, $\frac{1}{|b_2-\bp_3|}$ )
\\
(p) & max( $\sqrt{(1-x_1)(1-\xp_1)\rho}\Mb$, $\frac{1}{|b_2-b_q|}$ )
    & max( $\sqrt{x_3 \xp_3\rho}\Mb$, $\frac{1}{|\bp_1|}$ )
\\
(q) & max( $\sqrt{(1-x_1)(1-\xp_1)\rho}\Mb$, $\frac{1}{|\bp_2-\bp_3+b_3|}$)
    & max( $\sqrt{x_3 \xp_3\rho}\Mb$, $\frac{1}{|\bp_2-\bp_3|}$ )
\\
(r) & max( $\sqrt{(1-x_1)(1-\xp_1)\rho}\Mb$, $\frac{1}{|b_2|}$)
    & max( $\sqrt{x_3 \xp_3\rho}\Mb$, $\frac{1}{|\bp_1|}$ )
\\ \hline
\end{tabular}

\newpage

Table II: The coefficients $A_1$, $A_2$, $B_1$, and $B_2$ from the
transverse component $\phi^T$ of the $\Lambda$ baryon distribution
amplitudes for the choices of the hard scales,

(a) $t^i =\max\left(t^i_1,t^i_2,w,w',w_q\right)$,
(b) $t^i =\max\left(\frac{t^i_1+t^i_2}{2},w,w',w_q\right)$.
\vskip 0.5cm
\begin{tabular}{c|c|c}
\hline \hline
  &  (a)  & (b)  \\ \hline
$A_1^T$ & $-7.17\times 10^{-10}-8.69\times 10^{-9} i$
      & $-1.18\times 10^{-9}-1.54\times 10^{-8} i$
\\ \hline
$A_2^T$ & $-5.91\times 10^{-10}-1.62\times 10^{-8} i$
      & $5.97\times 10^{-10}-2.83\times 10^{-8} i$
\\ \hline
$B_1^T$ & $6.54\times 10^{-10}-7.53\times 10^{-9} i$
      & $1.14\times 10^{-9}-1.27\times 10^{-8} i$
\\ \hline
$B_2^T$ & $2.63\times 10^{-11}+1.21\times 10^{-8} i$
      & $6.36\times 10^{-10}+1.83\times 10^{-8} i$
\\ \hline
\end{tabular}

\vskip 1.0cm

Table III: The $\Lambda_b\to \Lambda J/\psi$ branching ratios and the
asymmetry parameters $\alpha$ derived in different approaches. The data
are also shown.
\vskip 1.0cm

\begin{tabular}{c|c|c|c|c|c}
\hline \hline
 & Ref. \cite{CT1} & Ref. \cite{FR} & Ref. \cite{MO} & This work &
Experimental data
\\ \hline
B($\Lambda_b\to \Lambda J/\psi$) & $2.1 \times 10^{-4}$ & $6.04
\times 10^{-4}$ & $2.49 \times 10^{-4}$ & $(1.65 \sim 5.27)
\times$ $10^{-4}$
 & $(4.7 \pm 2.8)\times 10^{-4}$
\\ \hline
$\alpha$ & -0.11 & -0.18 & -0.208 & $-0.17 \sim -0.14$ & -
\\ \hline
\end{tabular}

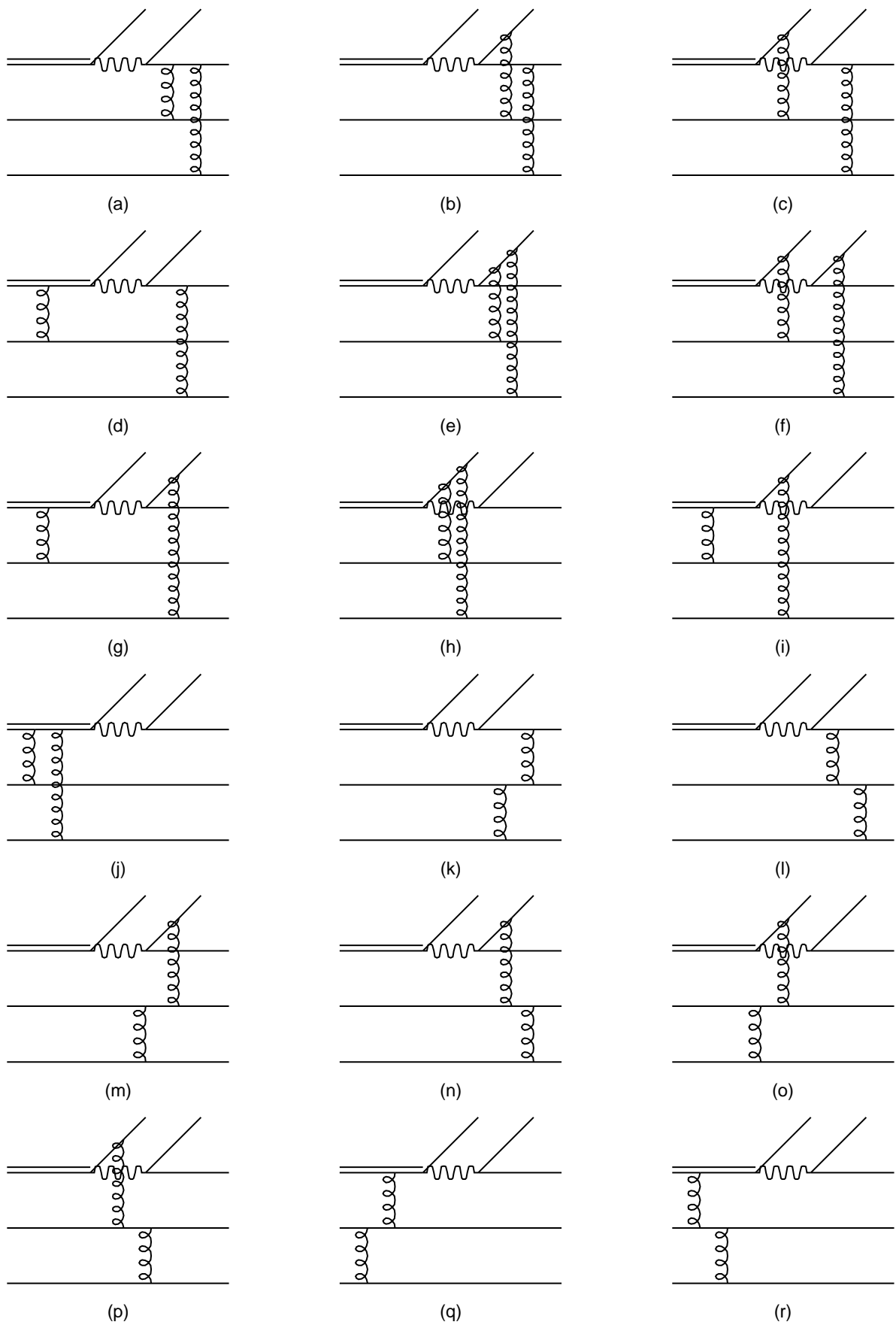
\end{document}